\documentclass[aps,prd,nofootinbib]{revtex4}

\usepackage{graphicx}
\usepackage{amsmath,amssymb,amsthm,amsfonts,amsbsy}

\def\x{{\boldsymbol x}}
\def\k{{\boldsymbol k}}
\def\y{{\boldsymbol y}}
\def\z{{\boldsymbol z}}
\def\p{{\boldsymbol p}}
\def\bs{\boldsymbol}

\newcommand{\ophi}{\overline{\phi}_0}
\newcommand{\ophic}{\overline{\phi}_c}

\begin{document}

%%%%%%%%%%%%%%%%%%%%%%%%%%%%%%%%%%%%%%%%%%%%%%%%
\title{\bf Quantum statistical correlations in thermal field theories:\\
boundary effective theory}

\author{A. {\sc Bessa}$^{1,2}$\footnote{abessa@ect.ufrn.br},
F. T. Brandt$^{2}$\footnote{fbrandt@if.usp.br},
C. A. A. {\sc de Carvalho}$^{3}$\footnote{aragao@if.ufrj.br}
and E. S. {\sc Fraga}$^{3}$\footnote{fraga@if.ufrj.br}}

\affiliation{$^{1}$Escola de Ci\^encias e Tecnologia, Universidade Federal do Rio Grande do Norte,
Caixa Postal 1524, 59072-970, Natal, RN, Brazil \\
$^{2}$Instituto de F\'\i sica, Universidade de S\~ao Paulo,
Caixa Postal 66318, 05315-970, S\~ao Paulo, SP , Brazil \\
$^{3}$Instituto de F\'\i sica, Universidade Federal do Rio de Janeiro,
Caixa Postal 68528, 21941-972, Rio de Janeiro, RJ , Brazil}

\date{\today}

%%%%%%%%%%%%%%%%%%%%%%%%%%%%%%%%%%%%%%%%%%%%%%%%

\begin{abstract}
We show that the one-loop effective action at finite temperature for a scalar field with quartic 
interaction has the same renormalized expression as at zero temperature if written in terms of 
a certain classical field $\phi_c$, and if we trade free propagators at zero temperature for their 
finite-temperature counterparts. The result follows if we write the partition function as an integral 
over field eigenstates (boundary fields) of the density matrix element in the functional 
Schr\"{o}dinger field-representation, and perform a semiclassical expansion in two steps: first, 
we integrate around the saddle-point for fixed boundary fields, which is the classical field 
$\phi_c$, a functional of the boundary fields; then, we perform a saddle-point integration over 
the boundary fields, whose correlations characterize the thermal properties of the system. 
This procedure provides a dimensionally-reduced effective theory for the thermal system. 
We calculate the two-point correlation as an example. 
\end{abstract}
%%%%%%%%%%%%%%%%%%%%%%%%%%%%%%%%%%%%%%%%%%%%%%%%
\maketitle

\vskip 5mm
%%%%%%%%%%%%%%%%%%%%%%%%%%%%%%%%%%%%%%%
\section{Introduction}

Effective actions are generating functionals for the one-particle irreducible vertices of a field theory. 
From the vertices, one obtains all the correlations of the theory, which may be used to compute 
physical quantities, after suitable renormalization. In finite-temperature field theories \cite{FTFT-books}, 
the effective action is the thermal equilibrium Gibbs free energy of the system of relativistic quanta of 
the theory. Effective actions are, therefore, natural quantities to compute at finite temperature. 
For instance, effective action calculations \cite{Blaizot:2000fc,Peshier:1998rz,Berges:2004hn,Berges:2005hc,
Arrizabalaga:2006hj,Borsanyi:2007bf} are among the techniques used to improve
 the poor convergence of the perturbative series in the bosonic sector, plagued by infrared divergencies.

In this paper we build an effective field theory for quantum statistical correlations using the boundary 
fields as the relevant degrees of freedom, which we denote by {\it boundary effective theory} (BET). 
To do so, we start from the partition function of our field theory in equilibrium at temperature $T=1/\beta$ 
written as a functional integral of the diagonal element of the density matrix (the Boltzmann operator for 
the field theory Hamiltonian) in the Schr\"{o}dinger field-representation. The integral is performed over 
the eigenvalues of the field operator in that representation, which are the stochastic variables of the 
problem, and which we will henceforth call boundary fields.

The density matrix element itself may also be written as a functional integral, one that describes an 
Euclidean time $\tau$ evolution from a given boundary field configuration at $\tau=0$, to that same 
configuration at $\tau=\beta$. The integral described in the previous paragraph is then performed 
over boundary fields. The quantum statistical correlations of the boundary fields should fully characterize thermal equilibrium.

The use of the density matrix functional leads to the derivation of the effective quantum statistical 
description for an underlying field theory, given by the effective action constructed from its microscopic 
Hamiltonian. The correlations of the stochastic fields (boundary fields, which only depend on spatial 
coordinates) express the thermalization encoded in the density matrix.

The density matrix functional was already used in a similar context in \cite{deCarvalho:2001xv}, 
where it led to the construction of dimensionally-reduced effective
actions. Likewise, in \cite{Bessa:2007vq}, it was used up to one-loop order
to investigate the thermodynamics of scalar fields. In the latter reference, the boundary
fields (considered as fluctuations around a homogeneous background)
played a very nontrivial role in the calculation of the partition function of scalar fields. 
In quantum mechanics, this formalism was developed in 
Refs. \cite{deCarvalho:1998mv,deCarvalho:1999fi,deCarvalho:2001vk} and led to the 
functional density matrix formulation of quantum statistics in Ref. \cite{Bessa:2008xj}.

Our goal, in the present article, is to use the density matrix functional in a scalar field theory with 
quartic interaction to demonstrate a simple relationship between effective actions at zero and 
finite temperature, computed up to one-loop order. The relation provides a natural bridge between 
correlations, as well as renormalization conditions, at zero and finite temperature. In practice, it 
yields a recipe to read off the renormalized finite temperature result from its renormalized 
zero-temperature counterpart.

The effective action that we obtain at finite temperature is a functional of the expectation value 
of the boundary (stochastic) fields. It is the generator of the one-particle irreducible vertices for the 
stochastic fields, and therefore allows us to reconstruct all stochastic correlations which define 
physical quantities in thermal equilibrium. In practice, our relation establishes a connection among 
those and zero-temperature physical quantities, such as particle masses and couplings.

This article is structured as follows: in Sec. \ref{densitymatrix}, we write both the density matrix and 
the partition function as functional integrals, define the generating functionals for quantum statistical 
correlations, and associate them to thermodynamic free energies; in Sec. \ref{fluctuations}, we 
integrate over the dynamical fields of the underlying theory, keeping the boundary fields fixed; in 
Sec. \ref{bcfluctuation}, we integrate over the (stochastic) boundary fields to obtain the effective 
action; in Sec. \ref{renormalization}, we go through the renormalization procedure to compute 
one-loop renormalized correlations; finally, in Sec. \ref{conclusions}, we present our conclusions.

%%%%%%%%%%%%%%%%%%%%%%%%%%%%%%%%%%%%%%%%%%%%%%%%
\section{Density matrix for scalar theories}
\label{densitymatrix}

In quantum statistical mechanics, the partition function for a system in contact with a thermal reservoir 
at temperature $T$ ($\beta=1/ T$) is expressed as a sum (integral) over a stochastic variable, whose 
probabilistic weight is given by the density matrix. For a system described by a (self-interacting) scalar 
field theory, the stochastic variable is the field eigenvalue in the functional Schrödinger field-representation, 
$\phi_0 (\x)$, which satisfies $\hat\phi|\phi_0 (\x)\rangle=\phi_0 (\x)|\phi_0 (\x)\rangle$.

The partition function is then a functional integral over the field eigenvalue of the diagonal element of 
the density matrix functional $\rho_{\beta}$,
%1
\begin{equation}
\label{Z1}
Z(\beta)=\int [{\cal D}\phi_0(\x)]\,\,\rho_{\beta}[\phi_0(\x),\phi_0(\x)]\;,
\end{equation}
%2
\begin{equation}
\label{phi}
\rho_{\beta}[\phi_0(\x),\phi_0(\x)]\equiv\langle\phi_0 (\x)|\exp(-\beta \hat{H})|\phi_0 (\x)\rangle\;.
\end{equation}
The Hamiltonian in the preceding formula specifies the underlying dynamics. In the present case, it is the dynamics of a (self-interacting) scalar field $\phi(\tau, \x)$.

As is well known \cite{Feynman-Hibbs}, $\rho_{\beta}$ can also be expressed as a functional integral  over dynamical fields $\phi(\tau, \x)$ defined for Euclidean time $\tau$, subject to the boundary conditions $\phi(0,\x)=\phi(\beta,\x)=\phi_0(\x)$,
%3
\begin{equation}
\label{rho}
\rho_{\beta}[\phi_0(\x),\phi_0(\x)]=\int\limits_{\phi(0,\x)=\phi(\beta,\x)=\phi_0(\x)}
[{\cal D}\phi(\tau, \x)]\, e^{-S[\phi(\tau, \x)]}\;.
\end{equation}
The boundary conditions relate the stochastic variable of the integral in Eq. \eqref{Z1} to the dynamical fields that are integrated over in Eq. \eqref{rho}. The Hamiltonian
%4
\begin{equation}
H[\Pi,\phi]= \frac{1}{2}\Pi^2 + \frac{1}{2}({\bf{\nabla}} \phi)^2 + \frac{m_0^2}{2}\, \phi^2 + U(\phi)\;,
\end{equation}
which involves the time-dependent field $\phi(\tau, \x)$ and the conjugate momentum $\Pi(\tau,\x)$, leads to the Euclidean action (we assume $d$ spatial dimensions),
%5
\begin{equation}
S[\phi]=\int_0^{\beta} (d^Dx)_{_E}\, {\cal L}[\phi]\;,
\end{equation}
%6
\begin{equation}
{\cal L}[\phi]=\frac{1}{2}(\partial_{\tau}\phi)^2+ \frac{1}{2}({\bf{\nabla}} \phi)^2 + \frac{m_0^2}{2}\, \phi^2 + U(\phi)\;.
\end{equation}
In this paper, we use the shorthands $x \equiv (\tau,\x)$ and $\int (d^Dx)_{_E} \equiv \int_0^\beta d\tau\,\int d^d\x$, with $D=d+1$.

The density matrix is a functional of $\phi_0(\x)$ only. We may write
\begin{align}\label{sd}
\rho_{\beta}[\phi_0(\x),\phi_0(\x)]\, = \, e^{-S_d[\phi_0(\x)]}\;,
\end{align}
$S_d$ being a certain temperature-dependent dimensionally-reduced action. The field $\phi_0(\x)$, the argument of $S_d$, depends only on the $d$ spatial coordinates; all the $\tau$ dependence of the original $(d+1)$-theory has been eliminated through the $\phi$ integration. The remaining integral over $\phi_0(\x)$, required to obtain the partition function, is unrestricted (except for the vacuum boundary conditions that must be imposed at spatial infinity).

The fields $\phi_0(\x)$ are the natural degrees of freedom of the reduced theory.  Any thermal observable can be constructed by integrating the appropriate functional of $\phi_0(\x)$ over the fields $\phi_0(\x)$  weighed with the corresponding diagonal element of the density matrix. The Euclidean time evolution can be viewed as an intermediate step which calculates the weights.

Before proceeding, notice that the density matrix is {\it not}, in general, of the form $e^{-\beta H_d}$ with $H_d$ being independent of $\beta$ as in ordinary quantum statistical mechanics; its $\beta$ dependence is far more complicated. This had already been pointed out in the analogous discussion of the transfer matrix carried out in Ref. \cite{wilsonkogut}. The density matrix provides a direct but alternative way of deriving a dimensionally-reduced theory.

In order to construct generating functionals for the dimensionally-reduced theory, we couple the field $\phi_0(\x)$ to an external current $j(\x)$, obtaining a modified action,
\begin{align}\label{I[phi0]}
I[\phi_0,j] = S_d[\phi_0] - \beta \int d^d\x \,j(\x)\phi_0(\x)\;.
\end{align}
The associated partition function,
\begin{align}\label{Z[j,alpha]}
Z[j] = \int[D\phi_0(\x)]\;e^{-I[\phi_0,j]}
\end{align}
is the generating functional of quantum statistical correlation functions. 

The quantum statistical connected correlation functions are obtained as functional derivatives of the (Helmholtz) free energy functional $F[j(\x)]$, defined as
\begin{align}
F[j] \,=\, -\frac{1}{\beta}\,\lim_{V \rightarrow \infty}\,\log Z[j] \;.
\end{align}
In particular, the expectation value of the field $\phi_0$ is given by
\begin{align}
\langle \phi_0 (\x)\rangle_j\, =\, -\frac{\delta F[j]}{\delta j(\x)} \;.
\end{align}
The label $j$ in $\langle \phi_0 \rangle_j$ is to stress that such an expectation value is a response of the system to an external current. In the sequel, we will drop the index $j$ to simplify the notation.

Another important quantity in the theory is the effective action $\Gamma$,
\begin{align}
\Gamma [\langle \phi_0(\x)\rangle ] = F[j(\x)] + \int d^d\x \,j(\x)\,\langle \phi_0(\x)\rangle\;,
\end{align}
the Legendre transform of $F[j]$. The argument of the effective action is the expectation value of the field, an intrinsic characteristic of the system. $\Gamma$ is the generating functional of the one-particle irreducible quantum statistical correlations. We shall use it to impose renormalization conditions that connect with the physical (zero-temperature) parameters of the underlying field theory, and to implement a renormalization procedure. From a thermodynamical viewpoint, this effective action is the Gibbs free energy functional, essentially the pressure of the system as a function of $\langle \phi_0(\x)\rangle$. In magnetic systems, for instance, the effective action gives the dependence of the pressure on the magnetization.

%%%%%%%%%%%%%%%%%%%%%%%%%%%%%%%%%%%%%%%%%%%%%%%%
\section{Fluctuations at fixed boundary}
\label{fluctuations}

The standard one-loop computation of the effective action for a position dependent background is performed in detail in Ref. \cite{amit} (Appendix $6-1$). It makes use of the steepest-descent method, and can be thought as the first term of a semiclassical series. In the present case, one has a double integration to perform, which will require an adaptation of the standard techniques.

The first integral over $\phi(\tau, \x)$, in Eq. \eqref{rho}, will be dominated by configurations in the vicinity of classical solutions $\phi_c$ satisfying
\begin{subequations}\label{eqmotion}
\begin{gather}
\square_{_E} \phi_c + m_0^2\, \phi_c + U^{\prime}(\phi_c) = 0\;,\\
\phi_c(0,\x) = \phi_c(\beta,\x) = \phi_0(\x)\;,
\end{gather}
\end{subequations}
where $\square_{_E} = -(\partial_\tau^2 + \bs \nabla^2)$ is the Euclidean D'Alembertian operator. Let us, for simplicity, assume that $U$ is a single-well potential. Then, a unique solution $\phi_c(\tau,\x)$ will exist for each boundary configuration $\phi_0(\x)$. The classical solution is a functional $\phi_c[\phi_0]$ of the boundary field. We write
\begin{gather}\nonumber
\phi(\tau,\x) = \phi_c(\tau,\x) +\eta(\tau,\x)\;, \\\label{decomposephic}
\eta(0,\x) = \eta(\beta,\x) = 0\;,
\end{gather}
and expand the action around $\phi_c$ to quadratic order in $\eta$,
\begin{align}
S[\phi] = S[\phi_c] + \delta^{(1)}S[\phi_c,\eta] + \delta^{(2)}S[\phi_c,\eta] + {\cal O}(\eta^3)\;.
\end{align}
The first-order variation $\delta^{(1)}S$ is given by
\begin{eqnarray}\label{eq:delta1S}
\delta^{(1)}S
=
\int (d^Dx)_{_E}\;
\frac{1}{2}\left[\square_{_E}\phi_c(x)+m_0^2\,\phi_c(x)+U^\prime(\phi_c(x))
\right] \eta(x)
+
\int d^d\x \left[\eta(x)\,\partial_{\tau}\phi_c(x)\right]_{0}^{\beta}\; ,
\end{eqnarray}
where we used the notation $\left[A(\tau)\right]_{0}^{\beta} = A(\beta)- A(0)$. The integrand in the first term of the right hand side of \eqref{eq:delta1S} vanishes
identically because $\phi_c(x)$ obeys \eqref{eqmotion}. The second term -- a
boundary term -- vanishes due to the property \eqref{decomposephic} of the fluctuation $\eta$.
Therefore, $\phi_c$ is an extremum of $S$ for $\eta$-like variations (fixed boundary).

The second-order variation $\delta^{(2)}S$ is
\begin{eqnarray}
&&
\delta^{(2)}S
=\frac{1}{2}\,
\int (d^Dx)_{_E}
\Big[\Big(\partial_\mu\eta(x)\Big)\Big(\partial^\mu\eta(x)\Big)
\!+\!
m_0^2\,\eta^2(x)
\!+\!
U^{\prime\prime}(\phi_c(x))\,\eta^2(x))\Big]
\nonumber\\
&&
\qquad=\frac{1}{2}\,
\int (d^Dx)_{_E}\;
\partial_\mu\Big[\eta(x)\,\partial^\mu\eta(x)\Big]
\nonumber\\
&&
\qquad\qquad
+\frac{1}{2}\,\int (d^Dx)_{_E}\;
\eta(x)\Big[\square_{_E}+m_0^2+U^{\prime\prime}(\phi_c(x))\Big]\,\eta(x)\; .
\end{eqnarray}
Using again \eqref{decomposephic}, we obtain, up to quadratic order in $\eta$,
\begin{eqnarray}
S[\phi] \approx S[\phi_c] + \frac{1}{2}\,
\int (d^Dx)_{_E}\;
\eta(x)\,\Big[\square_{_E} + m_0^2+ U^{\prime\prime}(\phi_c(\x))\Big]\,\eta(x)\;.
\end{eqnarray}
It is convenient to introduce the Green's function
\begin{subequations}\label{eq:prop}
\begin{eqnarray}
\left[\square_{_E} + m_0^2 + U^{\prime\prime}\left(\phi_c(x)\right)\right]
G[\phi_c](x;x^\prime)
=\delta^{(4)}(x-x^\prime)\\
G[\phi_c](\tau,\x;0,\x^\prime)=G[\phi_c](\tau,\x;\beta,\x^\prime)=0\;.
\end{eqnarray}
\end{subequations}
In terms of $G[\phi_c]$, we obtain
\begin{equation}\label{intovereta}
\int\limits_{\phi(0,\x)=\phi(\beta,\x)=\phi_0(\x)}\!\!\!\!\!\!\!\!\!\!\!\![D\phi(\tau,\x)]
\,e^{-S[\phi]}\; \approx \;e^{-S[\phi_c]}\; (\det G[\phi_c])^{1/2}\;.
\end{equation}
For single-well potentials it can be shown that $\delta^{(2)}S > 0$, a necessary condition for $\phi_c[\phi_0]$ to be a minimum  of $S$.

%%%%%%%%%%%%%%%%%%%%%%%%%%
\subsection{Building the classical solution}
\label{building}
In Appendix A, we obtain the following recursive relation for the classical solution $\phi_c[\phi_0]$,
\begin{eqnarray}
&&
\phi_c(\tau,\x)
=
\int d^d\x^\prime\;
\phi_0(\x^\prime)\;
\Big[
\partial_{\tau^\prime}G_0(\tau,\x;\tau^\prime,\x^\prime)
\Big]_{0}^{\beta}
\nonumber\\
&&
\qquad\qquad\qquad
-\int_{0}^{\beta}d\tau^\prime\int d^d\x^\prime\;
G_0(\tau,\x;\tau^\prime,\x^\prime)\;
U^\prime(\phi_c(\tau^\prime,\x^\prime))
\; ,
\label{eq:green1}
\end{eqnarray}
where $G_0(x,x^\prime)$ is a Green's function of the free operator
\begin{subequations}\label{eq:bound_prop}
\begin{gather}
\left[\square_{_E} +m_0^2\right]
G_0(\tau,\x;\tau^\prime,\x^\prime)
=\delta^{(4)}(x-x^\prime)\;, \\
G_0(\tau,\x;0,\x^\prime)
=
G_0(\tau,\x;\beta,\x^\prime)
=0\; .
\end{gather}
\end{subequations}
Diagrammatically, the classical solution $\phi_c(x)$ can be represented as the sum of all the tree diagrams terminated by the boundary field $\phi_0(\x)$. It is also demonstrated in Appendix A that
\begin{align}\label{jacobian}
\frac{\delta \phi_c(\tau,\x)}{\delta \phi_0(\y)} &= \left[\partial_{\tau^{\prime}} G[\phi_c](\tau,\x;\tau^{\prime},\y)\right ]_0^\beta\;.
\end{align}
From the previous equation, if one knows the classical solution $\ophic$ associated to some boundary configuration $\ophi$, the approximate classical solution associated to the following fluctuation of the boundary condition,
\begin{align}
\phi_0(\x) = \overline{\phi}_0(\x) +\xi(\x)\,
\end{align}
is given by
\begin{align}\label{phicofphi0}
\phi_c[\phi_0](x) = \ophic(x) + \int d^d\y\, \xi(\y)\,\left[\partial_{\tau^{\prime}} G[\ophic](x;\tau^{\prime},\y)\right ]_0^\beta\;+\;{\cal O}(\xi^2)\;,
\end{align}
as found in \cite{Bessa:2007vq}. Another useful relation is
\begin{align}\label{jacobian2}
\frac{\delta \left [\partial_\tau \phi_c(\tau,\x)\right ]_0^\beta}{\delta \phi_0(\y)} &= \left[\partial_{\tau}\partial_{\tau^{\prime}} G[\phi_c](\tau,\x;\tau^{\prime},\y)\right ]_0^\beta\;.
\end{align}

Finally, notice that $\phi_c[\phi_0]$ is, in general, a non-periodic function of $\tau$. As long as the only condition on the configurations that enter in the calculation of $Z[j]$ is to coincide at $\tau=0$ and $\beta$, non-periodic configurations are allowed.

%%%%%%%%%%%%%%%%%%%%%%%%%%%%%%%%%%%%%%%%%%%%%%%%
\section{Fluctuating the boundary field}
\label{bcfluctuation}

From the first functional integration (see Eq. \eqref{intovereta}) comes out an approximation for $S_d$ defined in \eqref{sd},
\begin{align}
S_{d}[\phi_0] = S[\phi_c[\phi_0]] + \frac{1}{2}\hbox{ Tr} \log G^{-1}[\phi_c[\phi_0]]\;.
\end{align}
$S_d$ works as the classical action of the theory. Using again the recipe to obtain the one-loop quantum effective action (now for the expectation value of the field $\phi_0(\x)$), we arrive at
\begin{align}\label{gammasd}
\beta\,\Gamma[\langle \phi_0 \rangle] &= S_{d}[\langle \phi_0\rangle] + \frac{1}{2} \hbox{ Tr} \log \delta^{(2)}S_d[\langle\phi_0 \rangle]\;,
\end{align}
where $\delta^{(2)}S_d[\langle\phi_0 \rangle]$ is a shorthand for the second-order variation of $S_d$ around $\langle \phi_0 \rangle$. However, to be consistent with the one-loop approximation, $\delta^{(2)}S_d[\langle\phi_0 \rangle]$ has to be replaced by $\delta^{(2)}S[\langle\phi_0 \rangle]$. In fact, the expression for $\Gamma$ as given in \eqref{gammasd} contains only part of the higher-order terms. Therefore, we are left with the following one-loop effective action,
\begin{align}\label{gammasd2}
\beta\,\Gamma [\langle \phi_0\rangle]&= S[\phi_c] + \frac{1}{2} \hbox{ Tr} \log G^{-1}[\phi_c] + \frac{1}{2} \hbox{ Tr} \log \left [\frac{\delta^2 S[\phi_c]}{\delta \langle \phi_0 \rangle^2}\right ]\;,
\end{align}
where $\phi_c$ in this case is $\phi_c[\langle \phi_0 \rangle]$\;.

It is convenient to integrate the classical action by parts, obtaining
\begin{align}\label{actionintparts}
S[\phi_c] &= \int (d^Dx)_{_E} \, \left (\frac{1}{2} \phi_c \square_{_E} \phi_c + \frac{1}{2}m_0^2\, \phi_c^2 + U(\phi_c)\right ) + \frac{1}{2}\int d^d\x \,\phi_0(\x)\,\left [\partial_\tau \phi_c (\tau,\x)\right]_{0}^{\beta}\;.
\end{align}
The presence of a boundary term is a direct consequence of the non-periodicity of $\phi_c[\phi_0]$, pointed out in the previous section. Such boundary contribution introduces nonstandard terms in the calculations.

At one-loop order, the expectation value of $\phi_0$ is given by the saddle-point of the modified action $I[\phi_0,j]$ (see Eq. \eqref{I[phi0]}). Let us call $\ophi$ such an extremal configuration. Fluctuating the boundary field as
\begin{align}\label{boundaryfluct}
\phi_0(\x) = \overline{\phi}_0(\x) + \xi(\x)\;,
\end{align}
and denoting $\phi_c[\ophi]$ by $\ophic$, the first-order variation of $I[\phi_0,j]$ is given by
\begin{align}\nonumber
\delta^{(1)} I[\phi_0,j] & = \int (d^Dx)_{_E} \,\left [ \square_{_E} \ophic(x) + m_0^2\, \ophic(x) + U^{\prime}(\ophic(x))\right ]\,(\phi_c(x)-\ophic(x))\\
& \;\;\;\;\;\;\;\;\;\;\;\;\;\;\;\;\; +\int d^d\x \,\xi(\x)\,\left \{ [\partial_\tau \ophic(\tau,\x) ]_0^\beta -j(\x) \right \}\;.\label{delta1I}
\end{align}
By construction, $\phi_c[\ophi]$ satisfies the equation of motion \eqref{eqmotion}, so that the first term on the r.h.s. of \eqref{delta1I} vanishes. From \eqref{delta1I}, we obtain the saddle-point condition \footnote{When $j=0$, condition \eqref{saddlej} finally implies that $\phi_c$ is a periodic function of $\tau$ with period $\beta$ (in fact, the trivial solution), but this property is not shared by classical solutions in the presence of nontrivial currents.} for $\ophi$:
\begin{align}\label{saddlej}
 [\partial_\tau \phi_c[\ophi](\tau,\x) ]_0^\beta = j(\x)\;.
\end{align}
Notice that Eq. \eqref{saddlej} is a condition on the time derivative of $\phi_c$. Therefore, whenever $j(\x) \neq 0$ (even for constant currents) the classical configuration $\phi_c[\ophi]$ will depend on $\tau$ (even in the free case). 

For the second-order variation, we obtain
\begin{align}
\delta^2 I[\phi_0]  = \delta^2 I^{(a)}[\phi_0]  + \delta^2 I^{(b)}[\phi_0]\;,
\end{align}
where
\begin{align}\nonumber
\delta^2 I^{(a)}[\phi_0] = \frac{1}{2}\int (d^Dx)_{_E} (d^Dx)_{_E}^{\prime} (d^Dz)_{_E}\, &\left[\partial_{\tau^{\prime}} G[\ophic](z;\tau^{\prime},\x)\right ]_0^\beta \xi(\x) \left [ \square_{_E} + m_0^2 + U^{\prime\prime}(\ophic(z))\right ]  \\
&\hspace{3cm}\times\, \left[\partial_{\tau^{\prime}} G[\ophic](z;\tau^{\prime},\x^{\prime})\right ]_0^\beta \xi(\x^{\prime})
\end{align}
and
\begin{align}\label{delta2I}
\delta^2 I^{(b)}[\phi_0] = \frac{1}{2}\int d^d\x\, d^d\x^{\prime}\, \xi(\x)\, \left [\partial_\tau\partial_{\tau^{\prime}}G[\ophic](\tau,\x;\tau^{\prime},\x^{\prime})\right]_0^\beta\,\xi(\x^{\prime})\;.
\end{align}
However, only the term $\delta^2 I^{(b)}[\phi_0]$ contributes, because
\begin{align}\label{dazero!}
 \left [ \square_{_E} + m_0^2 + U^{\prime\prime}(\ophic(z))\right ]  \left[\partial_{\tau^{\prime}} G[\ophic](z;\tau^{\prime},\x^{\prime})\right ]_0^\beta =0\;.
\end{align}
%^
Using that the second-order variation of $S[\phi_c[\phi_0]]$ and $I[\phi_0,j]$ coincide, we can write
\begin{align}\label{sndvariation}
\frac{\delta^2 S[\phi_c[\phi_0]]}{\delta \phi_0(\x)\delta \phi_0(\x^{\prime})}\bigg |_{\phi_0=\overline{\phi}_0}&= \left [\partial_\tau\partial_{\tau^{\prime}}G[\ophic](\tau,\x;\tau^{\prime},\x^{\prime})\right]_0^\beta\,.
\end{align}
We now have:
\begin{align}\label{aux:gammaA}
\beta\,\Gamma[\ophi] &= S[\ophic] + \frac{1}{2} \hbox{ Tr} \log  G^{-1}[\ophic] + \frac{1}{2} \hbox{ Tr} \log\left( \left [\partial_\tau\partial_{\tau^{\prime}}G[\ophic]\right ]_0^\beta\right)\,.
\end{align}
Using Eq. \eqref{eq:prop} written in terms of $G_0$ (see Eq. \eqref{eq:bound_prop}), we are led to
\begin{align}
\left [\partial_\tau\partial_{\tau^{\prime}}G[\ophic]\right ]_0^\beta & = \left[\partial_\tau\partial_{\tau^{\prime}}\left\{ G_0\left( 1+ G_0 U^{\prime\prime}(\ophic)\right)^{-1}\right\}\right]_0^\beta\;.
\end{align}
We may sum the series to obtain
\begin{align}
\left [\partial_\tau\partial_{\tau^{\prime}}G[\ophic]\right ]_0^\beta
&= \left[\partial_\tau\partial_{\tau^{\prime}} G_0\right]_0^\beta + \left[\partial_\tau G_0\right]_0^\beta U^{\prime\prime}(\ophic)G_0^{-1}G[\ophic]\left[\partial_{\tau^{\prime}} G_0\right]_0^\beta \,.
\end{align}
When we take the logarithm, we will have
\begin{align}
\log \left [\partial_\tau\partial_{\tau^{\prime}}G[\ophic]\right ]_0^\beta= \log {\cal G}_0 +  \log \left[1 + {\cal G}_0^{-1}\left(\partial_\tau G_0 U^{\prime\prime}(\ophic)G_0^{-1}G[\ophic]\partial_{\tau^{\prime}} G_0\right)\right],
\end{align}
where we have denoted $\left[\partial_\tau\partial_{\tau^{\prime}} G_0\right]_0^\beta$ as ${\cal G}_0$. Taking the trace and using that $\hbox{ Tr}\, (AB)^n = \hbox{ Tr}\, (BA)^n$ to reorganize the series, we obtain
\begin{align}
\hbox{ Tr} \log \left [\partial_\tau\partial_{\tau^{\prime}}G[\ophic]\right ]_0^\beta 
&=  \hbox{ Tr} \log {\cal G}_0 +  \hbox{ Tr}\log \left[1 + G_0^{-1}G[\ophic]\left (\partial_{\tau^{\prime}} G_0{\cal G}_0^{-1}\partial_\tau G_0 \right ) U^{\prime\prime}(\ophic)\right]\,.
\end{align}
We may write
\begin{align}
\log \left ( G^{-1}[\ophic]\right ) = \log \left (G_0^{-1} \right ) \, +\,\log \big (G^{-1}[\ophic]G_0\big)\;.
\end{align}
Therefore,
\begin{align}\nonumber
\hbox{ Tr} \log\left( G^{-1}[\ophic]\right) &+ \hbox{ Tr}\log\left( \left [\partial_\tau\partial_{\tau^{\prime}}G[\ophic]\right ]_0^\beta\right) = \hbox{ Tr}\log \left( G_0^{-1}\right ) + \hbox{ Tr}\log {\cal G}_0 \,+ \\\label{aux:nada}
&\;\;\;\;\;\;\;\;\;\;\;\;+\hbox{ Tr}\log \left \{1 + \left (G_0 + \partial_\tau G_0 {\cal G}_0^{-1}\partial_{\tau^{\prime}}G_0 \right) U^{\prime\prime}(\ophic) \right \}\;.
\end{align}
In Appendix B we show that the combination which appears multiplying $U^{\prime\prime}(\ophic)$ in \eqref{aux:nada} is the usual thermal propagator $\Delta_F$ which is given, in spatial Fourier space, by
\begin{gather}\label{aux:thermalprop}
\Delta_F(\k;\tau,\tau^\prime) = \frac{1}{2\omega_\k}\left [(1+n(\omega_\k))e^{-\omega_k|\tau-\tau^\prime|} + n(\omega_\k)e^{\omega_k|\tau-\tau^\prime|} \right ]\;,
\end{gather}
where $\omega_\k = \sqrt{\k^2+ m_0^2}$ and $n(\omega_\k)$ is the Bose-Einstein distribution. In terms of $\Delta_F$, one finally obtains
\begin{align}\label{Gammafinal}
\beta\,\Gamma[\ophi] &= S[\ophic] + \frac{1}{2}\hbox{ Tr} \log \left (G_0^{-1}\right ) + \frac{1}{2}\hbox{ Tr}\log {\cal G}_0  +\frac{1}{2} \hbox{ Tr}\log \left (1 + \Delta_F\,U^{\prime\prime}(\ophic) \right )\;.
\end{align}
Eq. \eqref{Gammafinal} can be further simplified to
\begin{align}\label{Gammafinalb}
\beta\,\Gamma[\ophi] &= S[\ophic] + \frac{1}{2}\hbox{ Tr} \log \left (G_\beta^{-1}[\ophic]\right )\;,
\end{align}
where, using that  $\hbox{ Tr} \log \left (G_0^{-1}\right ) + \hbox{ Tr}\log {\cal G}_0 = \hbox{ Tr} \log \Delta_F^{-1}$ (see Appendix B),
\begin{align}
G_\beta[\phi_c]^{-1} = \Delta_F^{-1} + U^{\prime\prime}(\phi_c)\;.
\end{align}
Notice that the dependence of $\Gamma$ on the physical field $\phi_0$ comes from the nontrivial functional dependence of the classical configuration $\phi_c$ on $\phi_0$. 

%%%%%%%%%%%%%%%%%%%%%%%%%%%%%%%%%%%%%%%%%%%%%%%%

\section{Renormalization procedure for the single-well quartic interaction (D=4)}
\label{renormalization}

The effective action in Eq. \eqref{Gammafinal} is written in terms of non-renormalized parameters. We conclude from the above calculation that $\Gamma$ has the usual zero-temperature expression when viewed as a function of the classical field $\phi_c$. This fact suggests that the renormalization procedure should follow the standard recipe. In this section, we verify that for the single-well quartic interaction in d = 3 spatial dimensions.

We start by introducing a cutoff $\Lambda$ for integrations over momenta and adding to $\Gamma$ couterterms in order to obtain a renormalized effective action,
\begin{align}\label{Gammafinalc}
\beta\,\Gamma_R[\ophi] &= \beta\, \Gamma[\ophi]  - \frac{C_1}{2}\,\int (d^4x)_{_E} \,\ophic^2(x) - \frac{C_2}{4}\,\int (d^4x)_{_E} \,\ophic^4(x) - \frac{C_3}{2}\,\int (d^4x)_{_E} \,\left (\partial_\mu \ophic\right)^2(x)\;.
\end{align}
Suitable choices of $C_1$, $C_2$ and $C_3$ should render $\Gamma_R$ finite. Renormalization conditions can be fixed using correlation functions of the $\ophic$ fields. Each choice of counterterms under that recipe will correspond to a definite set of renormalization conditions in terms of physical correlations involving $\phi_0$. That translation demands, in principle, the full knowledge of $\phi_c$ as a function of $\ophi$. 

 Let us call $v_0(\x)$ the saddle-point boundary configuration corresponding to $j(\x)\equiv 0$. Functional derivatives of $\Gamma_R$ with respect to $\ophi(\x)$ evaluated at $v_0(\x)$ lead to the $n$-point $1PI$ renormalized vertex function. For single-well potentials, one finds $v_0(\x) = 0$, and $\Gamma_R[\ophi]$ admits the expansion
\begin{align}
\Gamma_R[\ophi] = \sum_{n=1}^{\infty}\frac{1}{n!} \int \Gamma_R^{(n)}(\x_1,\ldots,\x_n)\,\ophi(\x_1)\,\ldots\,\ophi(\x_n)\, d^{3}\x_1\ldots d^{3}\x_n \;,
\end{align}
where
\begin{align}\label{gammaR}
\Gamma_R^{(n)}(\x_1,\ldots,\x_n) = \frac{\delta^{(n)}\Gamma_R}{\delta \ophi{}_1 \ldots \delta \ophi{}_n}\bigg |_{\ophi=0}\;.
\end{align}
One can also expand $\Gamma_R[\ophi]$ in powers of $\ophic[\ophi]$ (using that $\phi_c[0]=0$),
\begin{align}\label{npointusual}
  \Gamma_R[\ophi] = \sum_{n=1}^{\infty}\frac{1}{n!} \int \widetilde{\Gamma}_R^{(n)}(x_1,\ldots,x_n)\,\ophic(x_1)\,\ldots\,\ophic(x_n)\, d^{4}x_1\ldots d^{4}x_n \;,
\end{align}
where
\begin{align}\label{gammaRc}
\widetilde{\Gamma}_R^{(n)}(x_1,\ldots,x_n) = \frac{\delta^{(n)}\Gamma_R}{\delta \ophic{}_1 \ldots \delta \ophic{}_n}\bigg |_{\ophic=0}\;.
\end{align}

A standard calculation yields
\begin{align}
\frac{1}{(2\pi)^4\delta^{(4)}(p_1+p_2)}\widetilde{\Gamma}_R^{(2)}(p_1,p_2) = p_1^2 + m_0^2 + \frac{\lambda}{2}\hbox{ Tr} \Delta_F  - C_1 -C_3\,p_1^2 
\end{align}
and
\begin{align}
\frac{1}{(2\pi)^4\delta^{(4)}( \sum_{i}^4 p_i)}\widetilde{\Gamma}_R^{(4)}(p_1,p_2,p_3,p_4) = \lambda - \frac{\lambda^2}{2}I_1  -6C_2 \;,
\end{align}
where
\begin{align}
I_1 = \int^{\Lambda} \frac{d^4k}{(2\pi)^4}\left [\Delta_F(k)\Delta_F(k+p_3+p_4) +\Delta_F(k)\Delta_F(k+p_1+p_4) +\Delta_F(k)\Delta_F(k+p_3+p_1) \right ] \,.
\end{align}

For massive theories, we use the following zero-temperature renormalization conditions,
\begin{align}
&\frac{1}{(2\pi)^4\delta^{(4)}(0)}\widetilde{\Gamma}_R^{(2)}(0,0) = m_0^2\;,\\
&\frac{1}{(2\pi)^4\delta^{(4)}(0)}\frac{d\widetilde{\Gamma}_R^{(2)}}{dp_1^2}(0,0) = 1\;, \\
&\frac{1}{(2\pi)^4\delta^{(4)}(0)}\widetilde{\Gamma}_R^{(4)}(0,\ldots,0) = \lambda_R\,.
\end{align}
In this case, the counterterms are
\begin{align}\label{CT1}
C_1 = \frac{\lambda}{2}\int^{\Lambda}\frac{d^4k}{(2\pi)^4}\Delta_F^0(k)\;,\;\;\;\;\;\;\;C_2 = -\frac{\lambda^2}{4}\int^{\Lambda} \frac{d^4k}{(2\pi)^4}(\Delta_F^0(k))^2\;\;\;\;\;\;\;\hbox{and}\;\;\;\;\;C_3 =0\;,
\end{align}
where 
\begin{align}
\Delta_F^0(k) = \frac{1}{k^2 + m_0^2}
\end{align}
is the zero-temperature free propagator in 4-dimensional Euclidean Fourier space. 

Alternatively, the renormalization conditions can be expressed in terms of correlations of the fields $\ophi$. In principle, the general n-point function $\Gamma_R^{(n)}$ can be obtained from \eqref{npointusual} if one writes $\phi_c[\phi_0]$ in a power series of $\phi_0$, as given by the implicit relation \eqref{eq:green1}. In particular, $\Gamma_R^{(4)}$ is a complicated combination of $\tilde{\Gamma}_R^{(2)}$ and $\tilde{\Gamma}_R^{(4)}$. On the other hand, the 2-point function admits a simple expression,
\begin{align}
\frac{\beta}{(2\pi)^3\delta(\p_1+\p_2)}\,\Gamma_R^{(2)}(\p_1,\p_2) = {\cal G}_0(\p_1)
+ \left(\frac{\lambda}{2}\,\hbox{ Tr} \Delta_F-C_1- C_3 \p_1^2 \right)\,\int d\tau \,\left[\partial_{\tau^{\prime}} G_0(\tau,\p_1)\right ]_0^\beta\left[\partial_{\tau^{\prime}} G_0(\tau,-\p_1)\right ]_0^\beta\;.
\end{align}
In particular, using the counterterms given in Eq. \eqref{CT1}, we obtain, in Fourier space,
\begin{align}\nonumber
&\frac{\beta}{(2\pi)^3\delta(\p_1+\p_2)}\,\Gamma_R^{(2)}(\p_1,\p_2) =  2\omega_{\p_1}\,\tanh \beta \omega_{\p_1}/2 +\frac{\lambda_R}{24\beta}\,\frac{\tanh \beta \omega_{\p_1}/2}{\beta\omega_{\p_1}} \, \left (1+ \frac{\beta\omega_{\p_1}}{\sinh \beta \omega_{\p_1}}\right )\;.
\end{align}
When the zero-temperature theory is massless, we adopt the following set of renormalization conditions,

\begin{align}
&\frac{1}{(2\pi)^4\delta^{(4)}(0)}\widetilde{\Gamma}_R^{(2)}(\mu,-\mu) = \mu^2\;,\\
&\frac{1}{(2\pi)^4\delta^{(4)}(0)}\frac{d\widetilde{\Gamma}_R^{(2)}}{dp_1^2}(\mu,-\mu) = 1\;, \\
&\frac{1}{(2\pi)^4\delta^{(4)}(0)}\widetilde{\Gamma}_R^{(4)}(p_1,\ldots,p_i)\bigg |_{\mu} = \lambda_R\;,
\end{align}
where the arguments of the four-point function are such that $\sum_i p_i=0$, $p_i^2=\mu^2$ and $(p_i+p_j)^2=4\mu^2/3$ for $i\neq j$. The quantity $\mu$ is the renormalization scale. We obtain the counterterms
\begin{align}\label{CT}
C_1 = \frac{\lambda}{2}\int^{\Lambda}\frac{d^4k}{(2\pi)^4}\Delta_F^0(k)\;,\;\;\;\;\;\;\;C_2 = -\frac{\lambda^2}{4}\int^{\Lambda} \frac{d^4k}{(2\pi)^4}\Delta_F^0(k)\Delta_F^0(k+\mu)\;\;\;\;\;\;\;\hbox{and}\;\;\;\;\;C_3 =0\;.
\end{align}
The renormalized coupling constant $\lambda_R$ satisfies the following renormalization group equation,
\begin{align}
\mu \,\frac{d\lambda_R}{d\mu} = \frac{3\lambda_R^2}{16\pi^2}\;+\;{\cal O}(\lambda_R^3)\;.
\end{align}
As a consequence, $\lambda_R$ runs with the renormalization scale $\mu$ as
\begin{align}
\lambda_R(\mu) = \lambda_R(2\pi /\beta) + \frac{3\lambda_R^2(2\pi/\beta)}{16 \pi^2}\,\log\frac{\mu}{2\pi/\beta}\,+\,{\cal O}(\lambda_R^3)\;.% \frac{\lambda_R(2\pi/\beta)}{1-(3\lambda/16\pi^2)\log (\mu/2\pi T)}\;.
\end{align}

\begin{figure}[t!]
\begin{center}
\rotatebox{0}{%
     \resizebox{12cm}{!}{\includegraphics{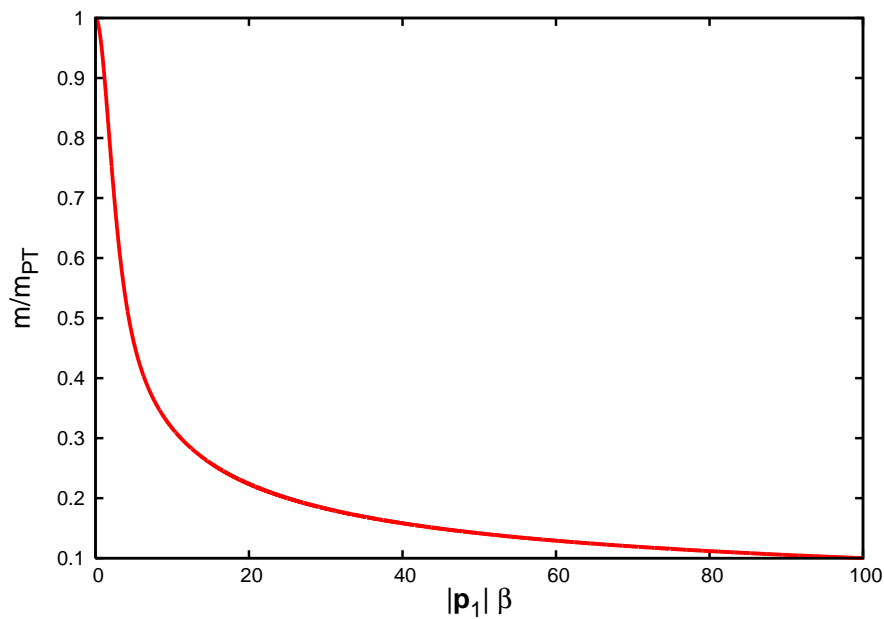}}}
\end{center}
\caption{\small Plot of the dynamical mass ($m$) of the boundary field $\phi_0$ (normalized by the first-order perturbative thermal mass $m_{PT}=\sqrt{\lambda_R/(24\beta^2)}$) as a function of $|\p_1|\, \beta$, where $\p_1$ is the external momentum. Notice that $m$ approaches $m_{PT}$ in the zero momentum limit.}\label{Fig:dynamass}
\end{figure}

The function $\Gamma_R^{(2)}$ in the massless case is given by
\begin{align}\label{aux:gamma2definem2}
&\frac{\beta}{(2\pi)^3\delta(\p_1+\p_2)}\,\Gamma_R^{(2)}(\p_1,\p_2;\mu) =  2|\p_1|\,\tanh \beta |\p_1|/2 +\frac{\lambda_R(\mu)}{24\beta}\,\frac{\tanh \beta |\p_1|/2}{\beta|\p_1|} \, \left (1+ \frac{\beta|\p_1|}{\sinh \beta |\p_1|}\right )\;.
\end{align}
The second term on the r.h.s. of \eqref{aux:gamma2definem2} plays the role of a mass term for the boundary theory (the first term is the kinetic part of that theory). 
The corresponding dynamical mass comes from the interaction of the boundary field with the $\eta$ fields and strongly depends on the external momentum, as shown in Fig. \ref{Fig:dynamass}.

One can think of an effective mass of the boundary field as the zero external momentum limit of the two-point function,
\begin{align}
\frac{1}{(2\pi)^3\delta(\p_1+\p_2)}\,\Gamma_R^{(2)}(0,0;\mu) \;=\; \frac{\lambda_R(\mu)}{24\beta^2}\;.
\end{align}
This effective mass coincides with the first-order perturbative result for the thermal mass ($m_{PT}$). The inverse of $\Gamma_R^{(2)}$ can be thought as the 3-dimensional propagator $G^{(2)}(\p_1,\p_2)$ of the field $\phi_0$. Numerical calculation shows the asymptotic behavior of $G^{(2)}(|x_1-x_2|)$ as
\begin{align}
G^{(2)}(|\x|) \propto \frac{\exp^{-m_{PT} |x|}}{|x|}\;,
\end{align}
which reinforces the role played by $m_{PT}$ as the mass scale for the boundary theory.

%

%%%%%%%%%%%%%%%%%%%%%%%%%%%%%%%%%%%%%%%%%%%%%%%%
\section{Conclusions}
\label{conclusions}

Naive perturbation theory applied to finite-temperature field theory is doomed to failure due to 
severe bosonic infrared divergences \cite{FTFT-books}. These are related to zero modes in the 
Matsubara frequencies, and introduce new mass scales associated with collective modes of the system. 
Incorporating this fact in the description of a thermal system in equilibrium naturally leads to the 
possibility of reorganizing the perturbative series, to different optimizations in the diagrammatic 
formulation, and ultimately to the construction of effective field 
theories \cite{Kraemmer:2003gd,Andersen:2004fp}. The process will be successful 
in the measure that one finds appropriate quasiparticles or degrees of freedom to anchor the 
effective theory. 

In this paper we have built an effective field theory that describes the quantum statistical correlations 
in thermal field theories using the expectation value of the boundary fields, which are the effective 
degrees of freedom, our natural stochastic variables and zero modes. 
This {\it boundary effective theory} (BET) is dimensionally-reduced by construction, as well as the 
thermal correlation functions it provides, which can be directly associated with the equilibrium 
thermal system.

To illustrate our formulation of the BET, we considered a scalar self-interacting field to one loop, 
but the procedure can in principle be extended to gauge fields and we believe that the result 
holds for higher loop orders. In the scalar case, we proved that the one-loop effective action at finite 
temperature has the same renormalized expression as at zero temperature provided it is written in 
terms of $\phi_{c}$ and zero-temperature propagators are substituted by their thermal versions. 
The classical solution $\phi_c(\tau,\x)$ is a functional of the boundary configuration $\phi_0(\x)$ and carries, in 
the underlying statistical mechanical problem, the dynamical information associated to the quantum functional integral 
formalism. Using that connection, one can compute directly all the renormalized thermal vertex functions and read off 
the renormalization group running of all parameters.

Even at one-loop order, the functional dependence of $\phi_c$ on $\phi_0$ 
led to a highly non-perturbative effective action for the boundary field. 
Our result for $\Gamma_R^{(2)}$ shows that, although the dynamical mass of the 
new degrees of freedom \---- the boundary fields \---- have a complicated momentum dependence, its 
zero external momentum limit coincides with the first-order perturbative result for the thermal 
mass and that $G^{(2)}$ has the correct asymptotic behavior as a function of $|x|$, an effect 
of dynamical screening. These are non-obvious consistency checks of the procedure. 

The rich structure encoded in the d-dimensional vertex functions obtained from the new 
degrees of freedom of BET will, of course, affect the thermodynamic functions and seem to 
provide a more direct and natural way to compute quantities such as the pressure. These 
results will be presented in a future publication \cite{future}.

%%%%%%%%%%%%%%%%%%%%%%%%%%%%%%%%%%%%%%%%%%%%%%%%
\begin{acknowledgments}

This work was partially supported by CAPES, CNPq, FAPERJ, FAPESP and FUJB/UFRJ.

\end{acknowledgments}

%%%%%%%%%%%%%%%%%%%%%%%%%%%%%%%%%%%%%%%%%%%%%%%%
\appendix

%%%%%%%%%%%%%%%%%%%%%%%%%%%%%%%%%%%%%%%%%%%%%%%%
\section{}

In this Appendix, we calculate the derivative of the functional $\phi_c[\phi_0]$. First of all, we multiply Eq. \eqref{eq:bound_prop}  by the classical field
$\phi_c(\tau^\prime,\x^\prime)$, and integrate over $\tau^\prime$ and
$\x^\prime$. This gives
\begin{equation}\label{phic=aux}
\phi_c(\tau,\x)=
\int_{0}^{\beta}d\tau^\prime\int d^d\x^\prime\;
\phi_c(\tau^\prime,\x^\prime)
\left[\square_{_E} +m_0^2\right]G_0(\tau,\x;\tau^\prime,\x^\prime)\; .
\end{equation}
Now, multiply the equation of motion (Eq. \eqref{eqmotion}) for
$\phi_c(\tau^\prime,\x^\prime)$ by the Green's function
$G_0(\tau,\x;\tau^\prime,\x^\prime)$ and integrate over $\tau^{\prime}$ and $\x^\prime$:
\begin{align}\label{eqmotiontimesG}
0 = \int_{0}^{\beta}d\tau^\prime\int d^d\x^\prime\,G_0(\tau,\x;\tau^{\prime},\x^{\prime})\,\left [(\square_{_E}+m_0^2)\phi_c(\tau,\x^{\prime}) + U^{\prime}(\phi_c(\tau^{\prime},\x^{\prime})) \right ]\;.
\end{align}
Subtracting \eqref{eqmotiontimesG} from \eqref{phic=aux} leads to
\begin{eqnarray}
&&
\phi_c(\tau,\x)
=
\int_{0}^{\beta}d\tau^\prime\int d^d\x^\prime\;
G_0(\tau,\x;\tau^\prime,\x^\prime)
\Big[
\stackrel{\leftarrow}{\square}_{_E}
-
\stackrel{\rightarrow}{\square}_{_E}
\Big]
\phi_c(\tau^\prime,\x^\prime)
\nonumber\\
&&
\quad
\;\;\;\;\;\;\;\;\;-\int_{0}^{\beta}d\tau^\prime\int d^d\x^\prime\;
G_0(\tau,\x;\tau^\prime,\x^\prime)\;
U^\prime(\phi_c(\tau^\prime,\x^\prime))
\; ,
\nonumber\\
&&
\label{eq:green}
\end{eqnarray}
where the arrows on the differential operators on the first line
indicate on which side they act. The first line can be rewritten as
a boundary term, by noting that
\begin{equation}
A\Big[\stackrel{\rightarrow}{\partial_\mu^2}
-
\stackrel{\leftarrow}{\partial_\mu^2}\Big]B
=
\partial^\mu \left\{
A\Big[\stackrel{\rightarrow}{\partial_\mu}
-
\stackrel{\leftarrow}{\partial_\mu}\Big]B
\right\}\; .
\end{equation}
In eq.~(\ref{eq:green}), the boundary in the spatial directions does
not contribute to the classical field at the point $\x$ because the
free propagator decreases fast enough when the spatial separation
increases. Thus, we are left with only a contribution from the
boundaries in time. At this point, since the boundary conditions for
$\phi_c$ consist in specifying the value of the field at
$\tau^\prime=0,\beta$, while its first time derivative is not
constrained, it is very natural to choose a Green's function $G_0$ that
obeys the following conditions,
\begin{equation}
G_0(\tau,\x;0,\x^\prime)
=
G_0(\tau,\x;\beta,\x^\prime)
=0\; .
\label{eq:bound_propb}
\end{equation}
With this choice of the propagator, we obtain formula
\eqref{eq:green1} for $\phi_c(\tau,\x)$. From \eqref{eq:green1}, we calculate
\begin{align}\nonumber
\frac{\delta \phi_c(\tau,\x)}{\delta \phi_0(\y)} &= \int d^d\x^{\prime}\,\delta(\x-\y)\,\left [\partial_{\tau^{\prime}}G_0(x;\tau^{\prime},\x^{\prime}) \right]_0^\beta \\
&\;\;\;\;\;\;\;\;\;\;\;\;\; -\int (d^Dx)_{_E}^{\prime}G_0(x,x^{\prime})U^{\prime\prime}(\phi_c(x^{\prime}))\,\frac{\delta \phi_c(\tau^{\prime},\x^{\prime\prime})}{\delta \phi_0(\y)}\\
&= \left [\partial_{\tau^{\prime}}G_0(x;\tau^\prime,\y) \right]_0^\beta - \int (d^Dx)_{_E}^{\prime}G_0(x,x^{\prime})U^{\prime\prime}(\phi_c(x^{\prime}))\,\frac{\delta \phi_c(\tau^{\prime},\x^{\prime\prime})}{\delta \phi_0(\y)}
\end{align}
Multiplying the previous equation by $G_0^{-1}(z,x)$ and integrating over $x$, we obtain
\begin{align}\nonumber
\int (d^Dx)_{_E}\, G_0^{-1}(z,x)\,\frac{\delta \phi_c(\tau,\x)}{\delta \phi_0(\y)} &=
 \int (d^Dx)_{_E}\, G_0^{-1}(z,x)\,\left [\partial_{\tau^{\prime}}G_0(x;\tau^\prime,\y) \right]_0^\beta \\
&\;\;\;\;\;\;\;- \int (d^Dx)_{_E} (d^Dx)_{_E}^{\prime} \,G_0^{-1}(z,x)G_0(x,x^{\prime})U^{\prime\prime}(\phi_c(x^{\prime}))\,\frac{\delta \phi_c(\tau^{\prime},\x^{\prime\prime})}{\delta \phi_0(\y)}\;.
\end{align}
Using that $G_0^{-1}(z,x) = \delta^{(4)}(z-x) \,(\square_{_E} + m_0^2)$, we can write
\begin{align}
(\square_{_E} + m_0^2)\,\frac{\delta \phi_c(z)}{\delta \phi_0(\y)} + U^{\prime\prime}(\phi_c(z))\frac{\delta \phi_c(z)}{\delta \phi_0(\y)} = \int (d^Dx)_{_E}\, G_0^{-1}(z,x)\,\left [\partial_{\tau ^{\prime}} G_0(\x,\tau ^{\prime};\y) \right]_0^\beta\;.
\end{align}
From \eqref{eq:prop}, we have
\begin{align}
G^{-1}[\phi_c](z^{\prime},z) = \delta^{(4)}(z^{\prime}-z)\,G_z^{-1}[\phi_c]\;,
\end{align}
where $G_z^{-1}[\phi_c] = \square_{_E} + m_0^2 + U^{\prime\prime}(\phi_c(z))$. Therefore,
\begin{align}
G^{-1}_z[\phi_c]\,\frac{\delta \phi_c(z)}{\delta \phi_0(\y)} = \int (d^Dx)_{_E}\, G_0^{-1}(z,x)\,\left [\partial_{\tau ^{\prime}} G_0(\x,\tau ^{\prime};\y) \right]_0^\beta\;.
\end{align}
Denoting $y$ by $(\tau^{\prime},\y)$ and $z$ by $(\tau^{\prime\prime},\z)$, notice that
\begin{align}
\partial_{\tau^{\prime}} \int (d^Dx)_{_E}\, G_0^{-1}(z,x)\,G_0(x;\tau^{\prime},\y) = \partial_{\tau^{\prime}} \left [ \delta^{(3)}(\z-\y) \delta(\tau^{\prime\prime} -\tau^{\prime})\right]_{0}^{\beta}\;,
\end{align}
so that
\begin{align}
G_z^{-1}[\phi_c]\,\frac{\delta \phi_c(z)}{\delta \phi_0(\y)} = \delta^{(3)}(\z-\y)\, \left [\partial_{\tau^{\prime}} \delta(\tau^{\prime\prime} -\tau^{\prime})\right]_0^{\beta}\;.
\end{align}
Multiplying both sides by $G[\phi_c](x,z^{\prime})\,\delta^{(4)}(z^{\prime}-z)$ and integrating over $(d^Dz)_{_E} (d^Dz)_{_E}^{\prime}$, we obtain
\begin{align}
\frac{\delta \phi_c(x)}{\delta \phi_0(\y)} &=  \int (d^Dz)_{_E} (d^Dz)_{_E}^{\prime}\,G[\phi_c](x,z^{\prime})G^{-1}[\phi_c](z^{\prime},z)\,\frac{\delta \phi_c(z)}{\delta \phi_0(\y)} \\
&= \int d\tau^{\prime\prime} d^d\z\, G[\phi_c](x;\tau^{\prime\prime},\z) \, \delta^{(3)}(\z-\y)\, \left [ \partial_{\tau^{\prime}} \delta(\tau^{\prime\prime} -\tau^{\prime})\right]_0^\beta\;\\
&= \int d\tau ^{\prime\prime} \,G[\phi_c](x;\tau^{\prime\prime},\y)\,\left [ \partial_{\tau^{\prime}} \delta(\tau^{\prime\prime} -\tau^{\prime})\right]_0^\beta\;.
\end{align}
Integrating by parts, Eq. \eqref{jacobian} follows.

%%%%%%%%%%%%%%%%%%%%%%%%%%%%%%%%%%%%%%%%%%%%%%%%
\section{}

The free propagator (solution of Eq. \eqref{eq:bound_prop}) can be explicitly calculated. It is given in Fourier space by
\begin{align}\label{eq:bound_prop2}
G_0(\tau,\tau^{\prime};\k) = \frac{\sinh [\omega_\k(\tau_{>}-\beta)]\,\sinh (\omega_\k\tau_{<})}{\omega_\k\,\sinh (\beta \omega_\k)}\;,
\end{align}
where $\tau_>(\tau_<) = $ max (min) $\{\tau,\tau^{\prime}\}$. From Eq. \eqref{eq:bound_prop2} we obtain
\begin{align}
\left [\partial_{\tau^{\prime}}G_0(\tau,\tau^{\prime};\k)\right ]_{0}^\beta = \frac{\cosh [\omega_\k(\beta/2-\tau)]}{\cosh (\omega_\k\beta/2)}
\end{align}
and
\begin{align}\label{fancyG0}
{\cal G}_0(\tau,\tau^{\prime};\k) =   \left [\partial_\tau\partial_{\tau^{\prime}}G_0(\tau,\tau^{\prime};\k)\right ]_{0}^\beta = 2\omega_\k \tanh (\omega_\k \beta /2)\;.
\end{align}
It is a simple matter to verify that
\begin{align}\nonumber
&\frac{1}{2\omega_\k}\left [(1+n(\omega_\k))e^{-\omega_\k|\tau-\tau^{\prime}|} + n(\omega_\k)e^{\omega_\k|\tau-\tau^{\prime}|}\right ] =  \frac{\sinh [\omega_\k(\tau_{>}-\beta)]\,\sinh (\omega_\k\tau_{<})}{\omega_\k\,\sinh (\beta \omega_\k)}\\
& \hspace{3cm}+  \frac{\cosh [\omega_\k(\beta/2-\tau)]}{\cosh (\omega_\k\beta/2)}  \frac{\coth \omega_\k \beta/2}{2\omega_\k} \frac{\cosh [\omega_\k(\beta/2-\tau^{\prime})]}{\cosh (\omega_\k\beta/2)}\;,
\end{align}
which is the expression in three dimensional Fourier space for
\begin{gather}
\Delta_F = G_0 + \partial_\tau G_0 \,{\cal G}_0^{-1}\,\partial_{\tau^{\prime}}G_0\;.
\end{gather}

It is possible to show that
\begin{align}
\hbox{ Tr} \log G_0^{-1}  \,=\, \int \frac{d^d\k}{(2\pi)^3} \, \log \frac{\sinh \beta \omega_\k}{ \omega_\k} \;.
\end{align}
Using \eqref{fancyG0} and the trigonometric identity
\begin{align}
2\tanh \frac{x}{2} \, \sinh x  = e^x \,(1-e^{-x})^2\;,
\end{align}
we obtain
\begin{align}
\frac{1}{2}\hbox{ Tr} \log G_0^{-1}  + \frac{1}{2}\hbox{ Tr} \log {\cal G}_0  \,&=\, \int \frac{d^d\k}{(2\pi)^3} \,\left [\frac{\beta \omega_\k}{2}  + \log (1-e^{-\beta\omega_\k})\right ]\\
&=\;\frac{1}{2} \hbox{ Tr} \log \Delta_F^{-1}  + C\;,
\end{align}
where $C$ is an unimportant infinity constant.

%%%%%%%%%%%%%%%%%%%%%%%%%%%%%%%%%%%%%%%

\end{document}